# QIS-XML: A metadata specification for Quantum Information Science

Pascal Heus, Richard Gomez
George Mason University, College of Science,  Department of Computational and Data Sciences
pascal.heus@gmail.com / rgomez@gmu.edu
December 2007

While Quantum Information Science (QIS) is still in its infancy, the ability for quantum based hardware or computers to communicate and integrate with their classical counterparts will be a major requirement towards their success. Little attention however has been paid to this aspect of QIS. To manage and exchange information between systems, today's classic Information Technology (IT) commonly uses the eXtensible Markup Language (XML) and its related tools. XML is composed of numerous specifications related to various fields of expertise. No such global specification however has been defined for quantum computers. QIS-XML is a proposed XML metadata specification for the description of fundamental components of QIS (gates & circuits) and a platform for the development of a hardware independent low level pseudo-code for quantum algorithms. This paper lays out the general characteristics of the QIS-XML specification and outlines practical applications through prototype use cases.

Keywords: quantum computer, information science, XML, interoperability, standards

# 1 Introduction

Quantum Information Science (QIS) is in its infancy and quantum computers are currently at an experimental stage. No production system exists and, while small prototypes are being tested in laboratories (using a few "qubits"), we may not see such reality for another 10 or 20 years. The challenge is that, while we do understand the underlying theoretical physics, the engineering needed to build quantum hardware is not yet here.

Nevertheless, quantum technology seem to be an inescapable outcome. At the current pace, Moore's law[1] will reach its limit around 2020[2]. Once transistors become a few or single atoms in size, quantum mechanical effects can no longer be ignored (today's typical transistors are as small as 20 atoms). While we do not know exactly when production grade quantum hardware will become available, rapid progress is being made and several prototypes have been demonstrated to work. Their existence is unavoidable given the physical limits of ever shrinking hardware.

QIS is not and cannot be a stand alone technology. The "Classic" Information Science (CIS) is already deeply rooted in our society and represents a significant segment of the global economy. This implies that to be successful, any kind of quantum information system will need to smoothly integrate in this existing environment. Interacting with quantum systems will also require an interface that will naturally be build on classic computers. This merging of technologies in essence can be seen as a unification of the two fields of expertise into a "Complete Information Science".

It is likely that the first generations of quantum computers will be  in the form of co-processors, remote systems or highly specialized circuits whose functionalities will simply extend standard computers and applications. Classic computers will remain the common interface for end-users and the development platform of choice for programmers (at least until we have a quantum operating system).

The classic information science will therefore need to become "quantum aware" to be able to manage QIS objects. This integration means that (1) classic and quantum computers will need to *communicate* with each other and (2) classic computers will need to be able to *represent and manage QIS concepts*.

A widely used technology that is particularly attractive to answer such requirements is the eXtensible Markup Language[3] or XML. Its primary purpose is to facilitate the sharing of structured data and metadata across different information systems. This paper introduces such approach through the proposal of QIS-XML, a metadata specification for Quantum Information Science.

---

1   http://en.wikipedia.org/wiki/Moore_Law
2   "Moore's Law is dead, says Gordon Moore", TechWorld, http://snipurl.com/1s2e7

---

3   http://en.wikipedia.org/wiki/Xml

# 2 Metadata and XML

Providing a comprehensive introduction to XML is outside the scope of this paper. A wide body of knowledge is available on the topic and numerous resources are available on the Internet. The key concepts are introduced below to facilitate the understanding of the paper.

## 2.1 Metadata

When we need to describe an object or an entity, we typically use adjectives to capture its fundamental characteristics. For example, if we want to describe a book, we can use its title, author, year of publication or ISBN number. For a car, a different set of attributes can be used such as manufacturer, brand, color, price, etc. In the information technology world, these characteristics are know as *metadata*[4] and is typically defined as *"data about the data"*. Metadata does not change anything about the item it describes but rather captures its nature by attaching a set of descriptive attributes to the object.

An important aspect of metadata is that it can be stored, accessed or exchanged with having to pass along the underlying object. We can browse a library catalog remotely or select a car from a brochure without going to the dealership.

## 2.2 The eXtensible Markup Language

Traditionally, when computer systems needed to capture such information, metadata was stored into a relational database system. While this works well in an isolated environment, databases run off proprietary solutions which make the exchange of information between agencies or individuals challenging. Database systems also are limited in functionalities and cannot always perform all the tasks we require for metadata management. The emergence of the Internet and of global communities has lead to new approaches based on openness and interoperability.

XML is one such technology. It allows computer systems to manage metadata (or data) in a harmonized way. It is non-proprietary (in the public domain) and can be used in any operating environment (Windows, Linux, Solaris, Mac). It is a language that "tags" or "markups" elements of information and stores them in a simple text format that can be read by any computer. XML and its related standards are maintained by the

World Wide Web[5] consortium (headed by by Sir Tim Berners-Lee[6], often referred to as the inventor of the World Wide Web).

XML is widely used on the Web, in particular to exchange information between organizations and different computer systems. It has emerged in recent years as a dominant technology. While invisible to most end users, XML is driving todays' Internet.

XML is actually a term that encompasses several technologies and functionalities. Adopting XML allows information systems not only to capture metadata (in XML) but also to validate it against agreed upon specifications (using DTD[7] or Schema[8]), transform it into other formats such as HTML or PDF (using XSLT[9]), search it like a database to lookup information (using XPath[10] or XQuery[11]), exchange it (using SOAP[12] or REST[13] based web services) and even edit it (using XForms[14]). All these functionalities are inherent to the XML technology and require little efforts to implement.

Here are simple examples of what an XML document looks like:

```
<catalog>
    <book isbn="0385504209">
        <title>Da Vinci Code</title>
        <author>Dan Brown</author>
    </book>
    <book isbn="0553294385">
        <title>I, robot</title>
        <author>Isaac Asimov</author>
    </book>
</catalog>

<mycontacts>
    <person>
        <name>John Doe</name>
        <email>jdoe@example.org</email>
    </person>
    <person>
        <name>Richard Feynman<name>
        <occupation>scientist</occupation>
    </person>
</mycontacts>
```

---

4   http://en.wikipedia.org/wiki/Metadata
5   http://www.w3.org/
6   http://en.wikipedia.org/wiki/Tim_Berners-Lee
7   http://en.wikipedia.org/wiki/Document_Type_Definition
8   http://en.wikipedia.org/wiki/XML_Schema
9   http://en.wikipedia.org/wiki/Xslt
10  http://en.wikipedia.org/wiki/Xpath
11  http://en.wikipedia.org/wiki/Xquery
12  http://en.wikipedia.org/wiki/SOAP
13  http://en.wikipedia.org/wiki/REST
14  http://en.wikipedia.org/wiki/Xforms

The language seem and is pretty simple. First it is human readable! XML is not a proprietary format and is stored as regular text. We can also see that "elements" of information are delimited by an opening and a closing "tag" (delimited by "<" and ">") that contains text or other elements (children). In some case, "attributes" can also be directly embedded in the tag itself as it is the case for the ISBN number of the book. While the syntax of XML contains a few more rules, this gives a general idea of its basic principles. What is important to understand is that it can be used to describe many many different things (it is extensible).

**Different metadata, different specifications**

Having a common language to describe data is a good start but is certainly not enough to satisfy all metadata management needs. Just like we create different databases structures, we also need a way to define the elements of information we want to use when describing an object (the semantic). For example, we would not describe a book or a car using the same attributes. Even if we are talking about the same object, it is very easy for different agencies or individuals to come up with different structures.

If we want to be able to universally manage and exchange metadata, we need to agree on common semantic to describe objects. This is what is known in XML as a "*specification*" and is defined using a Document Type Definition[15] or a an XML Schema[16]. The XML world is made of many specifications, each one specializing in a specific kind of object such as book, press release, car, weather, etc (the XML language itself is actually an XML specification). It is not uncommon for several specifications to exist for the same type of object and they sometime compete with each other. When this happens, it is then up to each agency to select one or to implement several at the same time (it is also actually quite easy to transform one specification into another).

XML specifications are typically maintained by an agency (or consortium) and made publicly available[17]. They can also be submitted to the International Standard Organization (ISO) to become an official ISO standard.

---

15 http://en.wikipedia.org/wiki/Document_Type_Definition
16 http://en.wikipedia.org/wiki/XML_schema
17 For example, see http://www.oasis-open.org or http://www.w3.org/

**XML and QIS**

When it comes to Quantum Information Science, no XML specification is readily available today. QIS however would greatly benefit from such standard as a common language to unify both fields of information science.

Given the wide acceptance of the technology amongst the Information Technology (IT) community, it would ensure that quantum based systems and related applications can dialog with their classic counterparts. It would also allow the development of harmonized tools as well as the integration of quantum functionalities into modern computer languages.

This paper introduces QIS-XML as a proposed XML specification for the description of fundamental components of QIS (gates & circuits) and a platform for the development of a hardware independent low level pseudo-code for quantum algorithms based on gates and circuits. Its basic concepts are introduced below along with a few use cases. More in depth information along with the schema and its documentation are available on the QIS-XML web site[18].

Note that the remaining of this document assumes familiarity with the basic concepts of quantum computing (qubits, gates, circuits, algorithms).

# 3 QIS-XML Model

The objective of QIS-XML is to provide a metadata model for representing quantum gates, quantum circuits and low level quantum pseudo-code. Creating such model in XML essentially consists in creating a schema that describes the various descriptive elements and attributes for each entity (the semantic).

When designing XML schema, various techniques are are commonly available. For QIS-XML, we choose a *venetian blind* design[19] to maximize component re-usability.

Note: to facilitate the comprehension of this section, we strongly recommend the reader to also consults the figures and illustrations available in annex.

## *3.1 Modules*

Instead of making QIS-XML a single large XML specification, we took a modular approach that allows users and developers to only use the components they need. This better fits in a distributed/federated environment where components can be designed by different agencies and referenced for reuse.

---

18 http://www.qisxml.org
19 http://www.xfront.com/GlobalVersusLocal.html

The QIS-XML specification is organized into the following modules:

– *Instance*: a global wrapper to bring together the other components (gates, circuits and programs);

– *Reusable*: describes a set of common elements that can be used by the all other modules;

– *Gate*: describes quantum gates and bring them together into gate libraries;

– *Circuit*: describes quantum circuits and bring them together into circuit libraries;

– *Program*: describes algorithms and bring them together into program libraries;

The first two modules exist primarily for technical reasons. The last three represent our quantum entities and build upon each other: gates provide the fundamental elements, circuits are made of gates, algorithms are made of circuits acting on quantum memory.

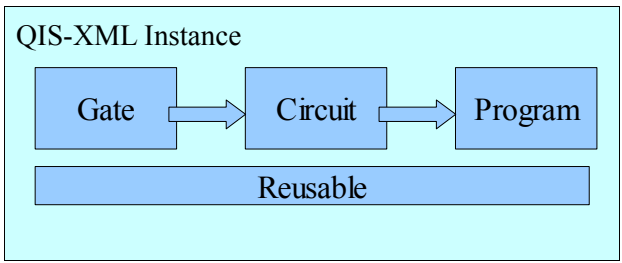

In XML, when designing a schema (such as a module), we associate it with what is called a *namespace*[20]. A namespace ensures that elements and attributes belonging to a schema do not conflict with another one carrying the same name in another schema. For example, the "title" element can exist in different schemas (a book, a person or a movie). In order to know which one we are talking about we need to define the context to which they belong.

The following namespaces have been defined for QIS-XML:

| Module | Namespace | Prefix |
|---|---|---|
| Instance | qis:instance:1_0 | i |
| Gate | qis:gate:1_0 | g |
| Circuit | qis:circuit:1_0 | c |
| Program | qis:program:1_0 | p |
| Reusable | qis:reusable:1_0 | r |

### 3.2 Instance Module

The instance module (Fig.1) is a wrapper and a very simple structure. It consists in a top level QIS element made of collections of gate, circuit or program libraries.

Note that libraries can be included directly in the document or by referencing external entities (the referencing mechanisms are explained below). The ability for the instance to point to external elements is particularly important as we do expect to bring together components from different places. For example, we could use the standard QIS-XML gate library, a few circuits designed by different people and write our own program. As these may be stored at different locations, an instance document allows to put these different pieces together. In nearly all cases, QIS-XML documents will be disseminated in the form of an Instance.

### 3.3 Reusable Module

The reusable module contains element types useful to all the other modules. It includes concepts fundamental to QIS as well as types to support referencing mechanisms.

**Fundamental types**

These elements either represent data types that do not readily exists in XML or are specific to QIS.

*Complex Number (Fig. 2)*: an element that can hold a *real* and *imaginary* value. As providing simple quantitative values is sometimes not sufficient, we also include the option to provide a *symbolic* expression that can be used to evaluate the value of a complex number.

*Matrix*: the Matrix type is defined as a *sequence of cells*. A Matrix comes with two attributes – @rows and @cols – to specify its dimensions. A Matrix Cell is then defined as a Complex Number with two extra

attributes – @row and @col – to specify its location of within the matrix. We made the general assumption that if a cell is not defined within a matrix, its value is zero.

*Qubit*: is simply defined as a collection of two complex numbers: Zero and One.

*Unitary Transformation (Fig. 2)*: is as a special type of matrix. Instead of having rows and columns, it is always square and has a @size attribute that specifies the number of qubits of the transformation. A size 1 is a 2x2 matrix, size 2 a 4x4, etc. It also comes with a Multiplier element as a complex number to scale the whole matrix.

**Example: Unitary Transformation for a Toffoli Gate**

```
<Transformation size="3">
    <Cell row="1" col="1" r="1"/>
    <Cell row="2" col="3" r="1"/>
    <Cell row="3" col="2" r="1"/>
    <Cell row="4" col="4" r="1"/>
    <Cell row="5" col="5" r="1"/>
    <Cell row="6" col="6" r="1"/>
    <Cell row="7" col="8" r="1"/>
    <Cell row="8" col="7" r="1"/>
</Transformation>
```

*Input* and *Output*: are simple structure containing a Name and a Description to provide extra information regarding a gate, circuit or program I/O.

**Referencing mechanism**

Given the modular approach of QIS-XML, we need a mechanism to point to various elements of information (they reside in different modules and can be stored at different locations). This requires two things: a way to *uniquely identify* an element and a way to make a *reference* to it.

You will shortly see that all elements that can be referenced (such as libraries, gates, circuits, memory) come with an *Identification* element (Fig.3). The identification is composed of a mandatory *ID* and optional elements to specify the agency maintaining the object and its version. The general rule is that an ID must be unique within its maintenance agency (this minimizes the risk of duplicate identifiers). Note that agency and version can be inherited from parents (a gate in a gate library by default belongs the the library's agency).

The other then complementary type is the *Reference* (Fig.3) that can be used to point to an identified element. It holds at least the element ID as well as an optional agency, version, and library ID (these are only really required in case of ambiguity). A URI attribute allows to make references to external documents (such as a library on the Internet or a local file).

**Example: Identification of a C-NOT Gate**

```
<r:Identification>
    <r:ID>C-NOT</r:ID>
</r:Identification>
```

**Example: reference to the above C-NOT Gate**

```
<c:GateRef>
    <r:ID>C-NOT</r:ID>
</c:GateRef>
```

### 3.4 Gate Module

This module is used to describe Quantum Gates[21]. It defines at the top level a Gate Library element that holds a collection of one or more gates. The Gate element (Fig.4) then holds all the characteristics of a gate.

Fundamentally, a quantum *Gate* can simply be defined as a unitary *Transformation*. The size attribute of the unitary transformation determines the size of the Gate (its number of qubits). To provide further information about the gate, we added a elements such as a *Name* (required), *Nickname* and *Description* as well as *Input* and *Output* to describe the gate's I/O. As we expect for a gate to be reused elsewhere (like in a circuit), we also have a mandatory *Identification* element.

In some cases, gates can be parametrized (their unitary transformation is not constant). To manage this case, we added a repeatable *Parameter* element whose *Name* can be reused in symbolic expressions of the transformation cells' values.

The Gate type also include a couple of extra elements. One is the *Image* element that can be used to provide a picture or illustration of the gate. Another one is *ProprietaryData* which allows any type of additional information for this specific gate. This has been included to allow manufacturers or applications to capture implementation specific information within QIS-XML.

**Example: Single qubit Z Gate**

```
<g:Gate>
    <r:Identification>
        <r:ID>Z</r:ID>
    </r:Identification>
    <g:Name>Pauli-Z</g:Name>
    <r:Transformation size="1">
        <r:Cell row="1" col="1" r="1"/>
```

---

21 http://en.wikipedia.org/wiki/Quantum_gate

```xml
            <r:Cell row="2" col="2" r="-1"/>
    </r:Transformation>
</g:Gate>
```

**Example: Single qubit Hadamard Gate**

```xml
<g:Gate>
    <r:Identification>
        <r:ID>H</r:ID>
    </r:Identification>
    <g:Name>Hadamard</g:Name>
     <r:Transformation size="1">
        <r:Multiplier r="0.707106781">
            <r:Symbolic
syntax="odf">1/sqrt(2)</r:Symbolic>
            <r:Symbolic
syntax="html">1/sqrt(2)</r:Symbolic>
        </r:Multiplier>
        <r:Cell row="1" col="1" r="1"/>
        <r:Cell row="1" col="2" r="1"/>
        <r:Cell row="2" col="1" r="1"/>
        <r:Cell row="2" col="2" r="-1"/>
    </r:Transformation>
</g:Gate>
```

**Example: Multi qubit Toffoli Gate**

```xml
<g:Gate>
    <r:Identification>
        <r:ID>TOFFOLI</r:ID>
    </r:Identification>
    <g:Name>Toffoli</g:Name>
    <g:Nickname>controlled-controlled-
not</g:Nickname>
    <g:Description>The Toffoli gate is a reversible
gate that takes three bits as input. The first two
are control bits and are left unchanged by the
gate. The third bit is flipped  if both control
bits are equal to 1. It is also known as the
"controlled-controlled-not" gate</g:Description>
    <r:Transformation size="3">
        <r:Cell row="1" col="1" r="1"/>
        <r:Cell row="2" col="3" r="1"/>
        <r:Cell row="3" col="2" r="1"/>
        <r:Cell row="4" col="4" r="1"/>
        <r:Cell row="5" col="5" r="1"/>
        <r:Cell row="6" col="6" r="1"/>
        <r:Cell row="7" col="8" r="1"/>
        <r:Cell row="8" col="7" r="1"/>
    </r:Transformation>
</g:Gate>s
```

Note that we do not expect to need many different gate libraries. The fundamental set of quantum gates is fairly well defined. At the times of this paper, the following gates have been described using QIS-XML:

– Single qubit gates: Hadamard, Identity, Pauli-X, Pauli-Y, Pauli-Z, Phase, Phase Shift, Square Root of Not, $\pi/8$
– Multiple qubit gates: Controlled-NOT (2), Controlled $\pi/8$ (2), Controlled Phase (2),

Controlled-Z (2), Deutsch Gate (3), Fredkin (3), Swap (2), Toffoli (3)

This library is publicly available and can be reused at will by referencing. It is however often easier to use a local verbatim copy in a local QIS instance to avoid having to resolve references.

### 3.5 Circuit Module

The circuit module (Fig.5) builds upon the gate module and introduce the notion of quantum circuits The module is defined at the top level as a Circuit Library element that holds a collection of one or more circuit or gate equivalence circuit (explained below).

Fundamentally, a quantum circuit has a specific *size* (it's number of input and output qubits) and is made of one or more *steps* that apply unitary transformation (gates) to its qubits. The QIS-XML model reflect this structure.

The *Circuit* element type (Fig.7) has been designed as a collection of *Step* elements made of one or more *Operation* elements. Steps represent the vertical cross-sections of a quantum circuit and an operation is used to *Map* the circuit's qubits to the input of a quantum *Gate* (or possibly another Circuit). Identification of a operation's gate (circuit) is made by reference. The operation element also provides the option to make a *Measurement* on a qubit. Like for gates, a few additional elements are available to provide descriptive information for the circuit: Name (required), Description, Input, Output, ProprietaryData along with a mandatory *Identification*.

**Example: 3 qubit phase flip circuit**

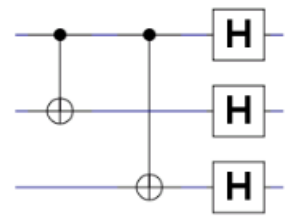

```xml
<c:Circuit size="3">
    <r:Identification>
        <r:ID>three_qb_phase_flip</r:ID>
    </r:Identification>
    <c:Name>3-qubit phase flip code</c:Name>
    <c:Step>
        <c:Operation>
            <c:Map qubit="1" input="1"/>
            <c:Map qubit="2" input="2"/>
            <c:GateRef>
                <r:ID>C-NOT</r:ID>
```

```xml
        </c:GateRef>
      </c:Operation>
    </c:Step>
    <c:Step>
      <c:Operation>
        <c:Map qubit="1" input="1"/>
        <c:Map qubit="3" input="2"/>
        <c:GateRef>
          <r:ID>C-NOT</r:ID>
        </c:GateRef>
      </c:Operation>
    </c:Step>
    <c:Step>
      <c:Operation>
        <c:Map qubit="1" input="1"/>
        <c:GateRef>
          <r:ID>H</r:ID>
        </c:GateRef>
      </c:Operation>
      <c:Operation>
        <c:Map qubit="2" input="1"/>
        <c:GateRef>
          <r:ID>H</r:ID>
        </c:GateRef>
      </c:Operation>
      <c:Operation>
        <c:Map qubit="3" input="1"/>
        <c:GateRef>
          <r:ID>H</r:ID>
        </c:GateRef>
      </c:Operation>
    </c:Step>
  </c:Circuit>
```

A few attributes have been included to meet special cases. For example, the map element can optionally specify a @value attribute to set the input to a fixed value of zero or one. A @reverse attribute is also attached to the operation element to signal that the associated unitary transformation should actually be reversed (apply the conjugate transform).

## Gate Equivalent Circuit

One of the functionality that we wanted to support in QIS-XML was the ability to describe "Gate Equivalent Circuits". This is motivated by the idea that any quantum gate can be represented in terms of sets of universal gates. For example, the set {Hadamard, Phase, C-NOT, pi/8} can be used to construct any quantum operation[22]. Other such examples include the 3-qubit Deutsch Gate or universal circuit as recently suggested by De Sousa and Ramos[23].

---

22 Quantum Computation and Quantum Information", Michael A. Nielsen, Isaac L. Chuang, 2000, p.189

23 "Universal quantum circuit for n-qubit quantum gate: A programmable quantum", P. B. M. Sousa, R. V. Ramos, 2006

The purpose of this *GateEquivalentCircuit* (Fig.6) element is to capture such information in anticipation that actual hardware implementation of quantum computers are likely to be limited to a specific set of universal gates. This means that a circuit described in QIS-XML is terms of "logical" gates will need to be transformed into a new circuit using the actual universal set supported by the hardware. Such transformation could be one of the steps performed by a proprietary quantum compiler. By including in QIS-XML the description of gates in terms of universal sets, it provides the ability to publish proprietary circuit equivalent libraries in a standard XML format.

The main difference between the *GateEquivalentCircuit* (Fig.6) and the regular circuit is that it adds a reference to the gate it replaces and includes the ability to remap the gate inputs to the circuit inputs (or fix their value). An optional *Model* element can also be used to identify a universal set or a specific manufacturer's computer model.

**Example: A SWAP gate equivalent using 3 C-NOT gates**

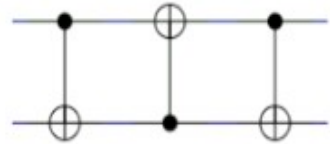

```xml
<c:GateEquivalentCircuit>
  <c:GateReference>
    <r:ID>SWAP</r:ID>
  </c:GateReference>
  <c:Circuit size="2">
    <c:Description>A SWAP gate equivalent using
3 C-NOT gates. The I/O are swapped as input of the
second gate.</c:Description>
    <c:Step>
      <c:Operation>
        <c:Map qubit="1" input="1"/>
        <c:Map qubit="2" input="2"/>
        <c:GateRef>
          <r:ID>C-NOT</r:ID>
        </c:GateRef>
      </c:Operation>
    </c:Step>
    <c:Step>
      <c:Operation>
        <c:Map qubit="1" input="2"/>
        <c:Map qubit="2" input="1"/>
        <c:GateRef>
          <r:ID>C-NOT</r:ID>
        </c:GateRef>
      </c:Operation>
```

```
        </c:Step>
        <c:Step>
            <c:Operation>
                <c:Map qubit="1" input="1"/>
                <c:Map qubit="2" input="2"/>
                <c:GateRef>
                    <r:ID>C-NOT</r:ID>
                </c:GateRef>
            </c:Operation>
        </c:Step>
    </c:Circuit>
</c:GateEquivalentCircuit>
```

### 3.6 Program Module

The program modules aims at the description of simple low level quantum algorithms. The module is defined at the top level as a Program Library element that holds a collection of one or more programs.

A *Program* (Fig.7) brings together two entities: the *algorithm* to be executed and the *quantum memory* it acts upon. The memory is a collection of qubits that can be organized in various *registers*. The algorithm is a collection of circuit *executions* and *measurements* (with the simple case being a single circuit and a single measurement).

#### The memory model

The initial design of the memory element was a simple quantum register but it was later influenced by Bernhard ¨Omer QCL[24] programming language whose memory management is very flexible. In QCL, a global quantum memory of N qubits is available and multiple quantum registers acting on a subset of the available qubit are defined (equivalent of variables). A similar model was also presented by Bettelli, S., L. Serafini, and T. Calarco in 2001. [25]. This same approach has been adopted in QIS-XML where we define a *Memory* (Fig.8) and a *Register* (Fig.9) type.

The only required element for *Memory* is its "size" attribute in number of qubits. Like most elements in QIS-XML, it also contains an Identification and a Name. A Memory also comes with a Prepare element (further described under Register below) and a collection of Qubit elements that can actually be used to store values if needed. Note that the Qubit element here is an extended version of the reusable r:Qubit types and adds an @index attribute to capture the memory location (1-based index).

---

24 http://tph.tuwien.ac.at/~oemer/qcl.html
25 http://arxiv.org/abs/cs/0103009

The *Register* is conceptually similar to a variable and can be defined globally (as a child of the program element) or within an Execute element (where it is mandatory). A Register is simply a subset of the program's Memory. It has a mandatory @size attribute in qubits and a set of elements used to describe which qubits of the main Memory it refers to.

**Example: A 6-qubit memory**
```
<p:Memory size="6"/>
```

**Example: A register using the whole memory with qubits 1 & 2 initialized to the value 1**
```
<p:Register size="6">
    <p:Prepare>
        <p:QubitSet>
            <p:QubitIndex>1</p:QubitIndex>
            <p:QubitIndex>2</p:QubitIndex>
            <p:Value r="1"/>
        </p:QubitSet>
    s</p:Prepare>
</p:Register>
```

There are three ways a register can make references to segment of the main program Memory:

–   The *QubitIndex* element can be used to refer to a specific qubit

–   The *QubitRange* (Fig.10) element can be used to refer to a range of of qubit using its *StartQubit* and *EndQubit* children elements

–   The *RegisterReference* can be used to refer to another Register definition

Note that combinations of the above three elements are allowed and that all index references in QIS-XML are 1-based (the first qubit is qubit 1, not 0). QIS-XML also makes the assumption that if no index or range is specified, it addresses the whole Memory starting from qubit 1 up to the @size of the register.

To set values of qubits in the register, we use the *Prepare* (Fig.11) element.

*Prepare* is a collection of one or more *QubitSet* (Fig. 10) elements that can be used to set the Value of specific qubit of the Register. It uses the same referential mechanism as for Memory except that here the index values are in relation to the Register (not the memory). The *Value* element is a complex number.

#### Algorithm: Execution and Measurements

The program itself or algorithm is made of a sequence of *Execute* (Fig.12) and *Measure* elements.

The Execute element is where the memory and circuits come together. Its basic version is composed of a Register element and a Circuit. The mandatory Register (who must have the same size of the Circuit) points to the qubits in Memory that the Circuit is acting upon. The *Circuit* itself can then be described directly (inline) as a child element or simply be a reference to a existing Circuit defined elsewhere in a Circuit library.

An experimental Program (or ProgramRef) element as also been included in the schema as in theory an Execute step could call a subprogram. Memory management becomes tricky in this case and this option needs further evaluation.

The *Measure* element in a program (not to be confused in Measurement under Circuit) can be used to specify qubit measurements after of between Execute steps. By default, if no Measurement element is specified, QIS-XML assumes that all qubits are measured and returned at the end of the Program execution.

Measurement can be used in two cases:

1. Specify which Memory qubit should be measure at the end of the Program. This is is desirable to reduce the number of required measurements or to only measure the relevant qubits (error correction or ancillary qubits can be ignored and may not be of interest)

2. If a program is composed of multiple Execute steps, it might be desirable to perform measurements between the Execute elements.

**A Program example**

The example below is a simple program that performs a 2+1 operation using a 2-qubit adder circuit (defined elsewhere in the demo library). This requires a 6-qubit Memory and contains a single execute step. The circuit's qubit 1 and 4 are used to specify the first number and qubit 2 and 5 the second. We therefore set qubit 4 and qubit 2 to specify the decimal number 2 and 1 respectively. Details on this particular adder circuit are presented later in this document.

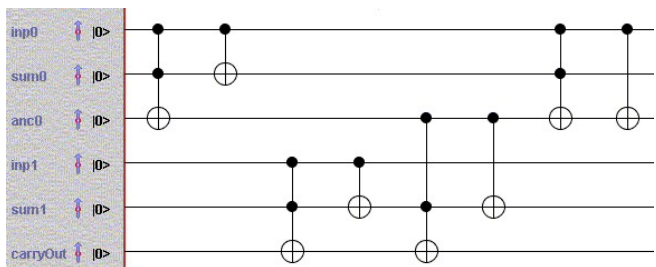

**Example: Performing the 2+1 operation using a 2-qubit adder circuit defined in a circuit library**

```
<p:Program>
    <r:Identification>
        <r:ID>two_plus_one</r:ID>
    </r:Identification>
    <p:Name>Two plus One</p:Name>
    <p:Memory size="6"/>
    <p:Execute>
        <p:Register size="6">
            <p:Prepare>
                <p:QubitSet>
                    <p:QubitIndex>2</p:QubitIndex>
                    <p:QubitIndex>4</p:QubitIndex>
                    <p:Value r="1"/>
                </p:QubitSet>
            </p:Prepare>
        </p:Register>
        <p:CircuitRef>
            <r:ID>adder2</r:ID>
        </p:CircuitRef>
    </p:Execute>
</p:Program>
```

This is a basic example but it illustrates the simplicity of QIS-XML

### 3.7 *Status*

This concludes the definition of QIS-XML. The schema and technical documentation are publicly available on the QIS-XML web site. While this current version is certainly usable (as we will demonstrate in the use case below), it should be considered a prototype. The specification need to be tested on complex use cases and will likely need adjustments, in particular the Program module. The Gate and Circuit modules can be considered fairly stable.

# 4 QIS-XML Transformation and Validation

We now have the ability to describe quantum gates, circuits and simple algorithms in a language that can be understood by all modern classic computers.

This sections examines other two extra aspects of metadata. One concerns representation: while XML is in theory human readable, this is usually not exactly what we want to show to the end-users. How can XML be delivered in a user-friendly fashion? The other is about quality control: how can we ensure that an XML document contains the information it claims to hold and that it meets its semantic rules.

## 4.1 From XML to HTML

As mentioned in the introduction, the XML technology comes with tools to transform XML documents into other formats. This process uses what are called Extensible Stylesheet Language Transformations[26] (XSLT). It is very commonly used to convert an XML into HTML web pages but can also create PDF documents, convert into other XML formats or simply output text.

Like nearly everything in the XML world, an XSLT is written using the XML language and therefore saved as an XML document, typically with the ".xsl" or ".xslt" extension. Explaining how XSLT are written is outside the scope of this document. Numerous resources and tutorials are available on the Internet[27]. In the context of QIS, it is interesting to note that XSL is a Turing complete[28] language.

To create an HTML page out of an XML document (like a QIS-XML file), we basically need an XSLT that extract the relevant information from the document and describes how it fits into the HTML page layout (HTML is concerned with presentation, not content). As this basically depends on the metadata elements contained in the document, a XSL transformation is typically designed to work with a single XML specification. It however works with any document that validates against the specific schema. For example, a XSLT designed to work with QIS-XML should be able to transform any valid QIS-XML document.

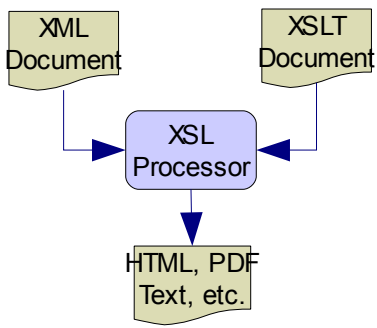

In order to perform the transformation itself, we need one more thing: an application that can combine the XML source document and the XSL Transformation to produce the actual output (see illustration). This is known as an XSL processor. XML being a global standard, such utility is actually built in most operating system and directly available in your favorite HTML browser. Most modern programming languages (C++, Java, .NET, pHp, etc.) also have the ability to natively perform such operation.

This means that if you have an XML document and a XSL transformation, you can basically use it on any computer without the need to install or develop any software. You can also integrate it into any application you might be developing. This begins to show some of the tremendous advantages of the XML technology: It works transparently across systems and is language independent.

With XSLT, we can now convert out test QIS-XML document that contains information (metadata) on gates and circuits into an HTML page. This basically is a *representation* of the XML content that can be displayed in a browser. There are of course many different "views" we can generate for our QIS-XML: list of gates, list of circuits, counting the number of gate and circuits, information on all the one qubit gates and circuits, etc. Each of these can be an individual XSLT or more complex parametrized transformations with reusable code. In the XML world, XSL is actually the closest thing to a programming language.

The examples provided in annex of this document have all been built using XSLT (Fig.13 & 14).

A useful concept to also introduce here as well is that an XML document is a well structured information container. An XSL transformation can "query" the content of the XML file to format it into an HTML for presentation to the user. This is another very powerful feature of XML: a document can be treated as a database system and therefore queried to extract information. Just like traditional databases have the Structured Query Language[29] (SQL) to perform searches, XML is equipped with a language called XPath[30] that can be used to retrieve metadata from a document. The major difference is that these queries can be performed by the XSL processor and therefore you do not need to install any new software on your computer or even design the database! It's build in the technology.

---

26 See http://en.wikipedia.org/wiki/XSL_Transformations
27 A good place to start is http://www.w3schools.com
28 See http://en.wikipedia.org/wiki/Turing_complete

29 See http://en.wikipedia.org/wiki/Sql
30 See http://en.wikipedia.org/wiki/Xpath

## 4.2 QIS-XML Validation

### XML and Schema validation

To ensure metadata quality, an important operation that can be performed natively by XML is called *validation*. This is typically a two step process. First an XML document can be checked if it is syntactically correct (does it follows the rules of the XML language?). If yes, a document is considered well-formed[31]. In such case, we can also further validate the document against it's schema. For example, the QIS-XML schema defined in the previous section defines a semantic made of elements and attributes that can be mandatory or repeatable. We can therefore verify that an XML document claiming to be in QIS-XML (by declaring membership in a QIS-XML namespace) meets these criteria. If so, the document is called valid[32].

Such validation can be performed by what is called an *XML parser*. A parser is a utility software that takes an XML document as input, controls its syntax against the rules of the XML language (like a spell checker) and then validates its content against the schema defined by the namespace(s). Both free open source and commercial parsers are widely available.

### Second level validation

XML Schema based validation however has its limits and cannot perform some of the advanced checks required by QIS-XML. For example, verifying that a qubit total probability is equal to one or that a circuit does not map the same qubit multiple time in a single step is beyond the semantic rules. Such second level validation can actually be achieved programmatically.

XSL Transformations introduced in the previous section for example can be used for such purpose as well. Fundamentally, performing validation consists in parsing the XML content to check its consistency and coherence and produce a report out of it. If the report happens to be in HTML format, it can be viewed in a browser. It is in essence the same as a transformation of the XML metadata. Simple XSLT based second level validation has been developed for QIS-XML and an output example is presented in Annex (Fig.15). We expect the validation mechanisms to grow as the QIS-XML increases in complexity.

### Validation benefits

While it is not always required to validate XML documents when working locally (known sources), it becomes an important feature when information is exchanged between various organizations to ensure that it meets the standard before being processed. Having such functionalities directly embedded in the technology minimizes the development and maintenance costs and facilitates quality assurance. These are all other good reasons to adopt an XML based framework.

# 5 QIS-XML Visualization

Another direct benefit of using QIS-XML is the ability to provide a graphical representation of gates, circuits and programs. Again, when you think about it, this is conceptually a transformation of the QIS-XML document into a diagram or an image. The natural XML reflex is therefore to think XSLT but the problem in this case is that image file are often proprietary binary formats (GIF, JPG, PNG, TIF, etc.). While XSLT could in theory produce such output, a specialized program develop using Java or C would be more efficient (and more appropriate).

Fortunately, XML comes again to the rescue. There is a W3C[33] XML specification called Scalable Vector Graphics[34] (SVG) that can be used to to describe two-dimensional vector graphics. Since a quantum circuit diagram can be broken down into a collection of lines, circle, boxes and text, we should therefore be able to "describe" the diagram as a composition of these basic elements.

Numerous resources are available on the Internet on SVG-XML[35]. While it provides a set of advanced functionalities, basic SVG is fairly straightforward to use. SVG can be displayed in most web browsers, either natively and after the installation of a free plugin[36].

---

31  http://en.wikipedia.org/wiki/Well-formed_XML_document
32  http://en.wikipedia.org/wiki/Well-formed_XML_document

33  See http://www.w3c.org
34  See http://en.wikipedia.org/wiki/Svg
35  A good place to start is http://www.w3schools.com/svg/default.asp
36  http://en.wikipedia.org/wiki/Svg#Support_for_SVG_in_web_browsers

## SVG using XSLT

Converting the QIS-XML into SVG-XML using pure XSL transforms is a bit of a challenging process. The representation of single qubit gates is simple but the automatic layout of a circuit diagram is more problematic. Issues such as connecting lines between gate's inputs/outputs, moving multi-qubit gate's nodes to the proper location, or fixed map @input value, required quite a lot of development effort. The implementation we have achieved works for circuits using well defined gates but will need improvements in order to manage more complex cases (such as circuits in circuit). As mentioned before, the XSLT language has its limitations. For complex cases, a more flexible programming language would be more appropriate

**Example:SVG representation of a Hadamard gate**

```
<svg:svg width="50" height="60">
  <svg:g transform="translate(25,25)">
    <svg:line x1="-25" y1="0" x2="-20" y2="0"
style="stroke:black;stroke-width:1;stroke-opacity:
1;"/>
    <svg:rect x="-20" y="-20" width="40"
height="40" style="opacity:1; fill:none;
stroke:black; stroke-width:1; stroke-opacity:1;"/>
    <svg:line x1="20" y1="0" x2="25" y2="0"
style="stroke:black;stroke-width:1;stroke-opacity:
1;"/>
    <svg:text x="-11" y="10"
style="font-size:30px; font-
weight:bold;">H</svg:text>
  </svg:g>
</svg:svg>
```

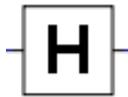

Additional examples are available in Annex (Fig.16 & 17).

## Benefits

Visualization is an important functionality when it comes to QIS. The ability to transform standard gates, circuits or even programs into a graphic or diagram representation is a powerful feature. While developing utility software that performs this with high efficiency may require some efforts, the benefits are considerable and again available natively in XML.

# 6 QIS-XML Applications

## 6.1 Gate / Circuit libraries

One of the advantages of QIS-XML is the ability to capture and exchange information about quantum circuits. The metadata exchange mechanisms are integrand part of the XML technology and allows developers to publish their circuits in global libraries and make them available to other users. By using a common language (QIS-XML), this sharing of information promotes a collaborative research and development environment. This is further facilitated by QIS-XML referencing mechanisms which allow users to remotely integrate circuits into their own designs or libraries. From the programming perspective, these can be seen as low level function libraries that can be called locally or remotely. The availability of standard visualization tools also facilitate the understanding of the circuits.

QIS-XML therefore provides a standard environment that can foster collaborative work and facilitate the understanding of the QIS technology.

## 6.2 Programming

The motivation being the programming module of QIS-XML is to provide quantum programming languages and quantum equipment manufacturers with a platform neutral low level pseudo-code. This is somewhat equivalent to a general purpose assembly language[37].

A quantum programming language compiler could for example convert high level code into low level QIS-XML program (using memory and circuits). This information can then sent to the actual quantum hardware. Such exchange process can be local (to a co-processor) or through quantum remote procedure calls (for example a service available on the Internet). We could even imagine distributing the code across multiple quantum processors, each specializing into its specific function.

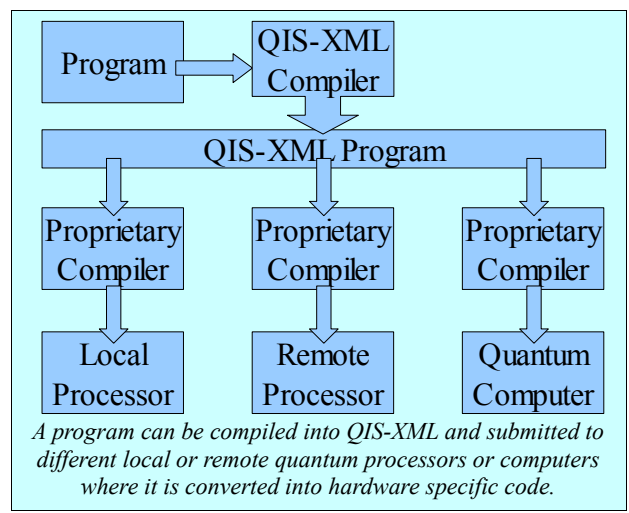

*A program can be compiled into QIS-XML and submitted to different local or remote quantum processors or computers where it is converted into hardware specific code.*

---

37 http://en.wikipedia.org/wiki/Assembly_language

The hardware itself then has a local compiler that turns QIS-XML instructions into hardware specific commands. This may involve converting logical gates into universal subsets implemented by the equipment (possibly described using QIS-XML equivalent circuits) and turning the XML into a proprietary set of instructions.

The advantages of this systems are that the high level language compiler does not need to be aware of the local hardware implementation of the circuits (or their equivalent) and the quantum processor can be located anywhere, the information exchange taking place using standard XML mechanisms (like SOAP[38] or REST[39]).

## Use case: Genadder & simulators

As a proof of concept for the programming framework, we used a simple quantum circuit generator and, through QIS-XML, converted its output to code that can be executed on various quantum computer simulators (given that we do not have hardware available today). This process is illustrated by Fig. 19.

The circuit generator we used is "Genadder"[40]. It was designed by NIST in 2003 to create quantum circuits to add two binary numbers of any width. The original version of Genadder outputs circuits description in a text format. We adjusted the C source code to instead generate the circuits in a QIS-XML format. Once this was completed, it allowed us to visualize the circuits using the QIS-XML to SVG-XML transforms (Fig.18) and to include the circuit in a QIS-XML program to perform the actual addition operation (Fig.20).

We then selected three quantum computer simulators to execute the program: the Fraunhofer Quantum Computing Simulator[41], the QCL[42] simulator and QuiDDPro[43] as target platforms. These were selected for their functionalities, stability, performance and product maturity. Many other simulators were identified but were no longer maintained, not well documented or inappropriate for our purpose (the complete list is available and maintained on the web[44]).

Each simulator using a different syntax, the next step consisted in converting the QIS-XML program into source code that could be executed on the selected simulators. This is a transformation of the XML into a text file and can be carried over using standard XSL Transformation. Like for SVG however, high performance converter would be better implemented in a common programing language such as Java or C. We did complete this step for the Fraunhofer (Fig.21) and the QCL simulator (Fig.22) but lack of time prevented us to generate code for the QuiDDPro simulator. We however achieved the expected results on the first two simulators and repeated the test for circuits with up to 50-qubits.

This simple example illustrate how QIS-XML could benefit both programmers and hardware implementers by having the high level code written in a common language, converted to QIS-XML circuits and programs, and transparently executed on different hardware platforms.

*Note that a detailed description of this process is available on the QIS-XML web site.*

## 6.3 Tools Interoperability

Another potential application of QIS-XML is its ability to provide a harmonized information storage framework for QIS tools in order to facilitate their interaction. Most of the QIS management and programming applications will actually be executed on classic computers and will need to exchange information with each other.

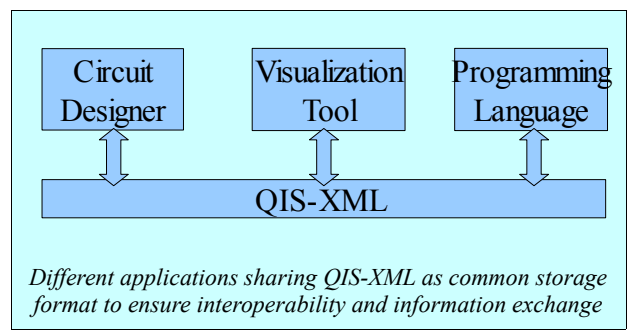

*Different applications sharing QIS-XML as common storage format to ensure interoperability and information exchange*

For example, a quantum circuit designer software could feed its output to a specialized visualization tool or function library to be used in a programming environment. Without a standard storage system, these applications are likely to use proprietary data formats and exchanging information would require complicated and unnecessary conversions. QIS-XML alleviate this issue by providing a common language for the capture of gate, circuits and low level programming instructions.

38 http://en.wikipedia.org/wiki/SOAP
39 http://en.wikipedia.org/wiki/REST
40 http://hissa.nist.gov/~black/Quantum/genadder.html
41 http://www.qc.fraunhofer.de/
42 http://tph.tuwien.ac.at/~oemer/qcl.html
43 http://vlsicad.eecs.umich.edu/Quantum/qp
44 http://rgomez.cos.gmu.edu/qis/?lvl1=simulators

# 7 Status and next steps

**Status**

While fully functional, the QIS-XML specification should currently be considered a prototype. It needs to be used by a broader range of users and will likely require several adjustments, in particular the programming module. Further testing will need to establish if the specification scales up to complex circuits or algorithms and is suitable as a neutral platform for tools and compilers. It can nevertheless be used today by anyone interested in applying or adopting the XML technology in the field of QIS. We hope this paper will incite others to experiment with it and contribute to its enhancement.

**Need for tools**

QIS-XML also need specific tools. The initial XSL transformations used to provide second level validation, generate the SVG, or convert programs to proprietary code have their limitations and could benefit from an implementation in a language such as Java or C. Utilities will also be required for the management of complex circuits and programs or to implement features such as external referencing.

**Hardware Device Capability Module?**

One potential extension to QIS-XML would be a quantum device capability module. One of the issue we identified while working on the code converter for the simulators is that we do not really know in advance what a hardware quantum component (or a software simulator) is capable of. This makes it difficult to establish whether it supports the execution of specific QIS-XML code. Having a standard structure to describe the characteristics of quantum hardware would be very useful to address this issue. Such module could capture information such as the number of qubits , type of gates supported by the component, the circuit equivalence model, ability to reverse gates, the memory system, the level of programmability, etc. Such module is a bit comparable to the device capability system commonly used by printers to describe their characteristics.

**Adoption & Sustainability**

Finally, like any XML specification, QIS-XML cannot be driven by a small group of individuals. To ensure global adoption, it will need to be endorsed at the institutional level by agencies and organizations who will support its development, foster its adoption, and make it a sustainable standard. This specific issue will actually be addressed in another paper suggesting the establishment of a QIS Alliance whose overall objective will be to examine the issues of integration and interoperability between classic and quantum information science. One of the activity of the QIS Alliance would be to maintain QIS-XML.

# 8 Conclusions

We have demonstrated through this work how the widely used XML standard could benefit quantum and classic information sciences by bridging both worlds. We also introduced QIS-XML as a potential solution to address fundamental information exchange and quantum programming issues.

We believe that the above is likely just scratching the surface of what is potentially achievable using QIS-XML and other QIS oriented other XML specifications.

We hope this initial effort will inspire others to look at quantum and classic information sciences as a unified science and to further investigate the issues of integration and interoperability. Classic computers will be around for a long time and will remain the preferred interface for end users and developers. The ability for quantum computing technology to communicate with classic computers along with the availability of relevant open tools will play a major role in the success and future of QIS.

# Annex 1: QIS-XML Objects

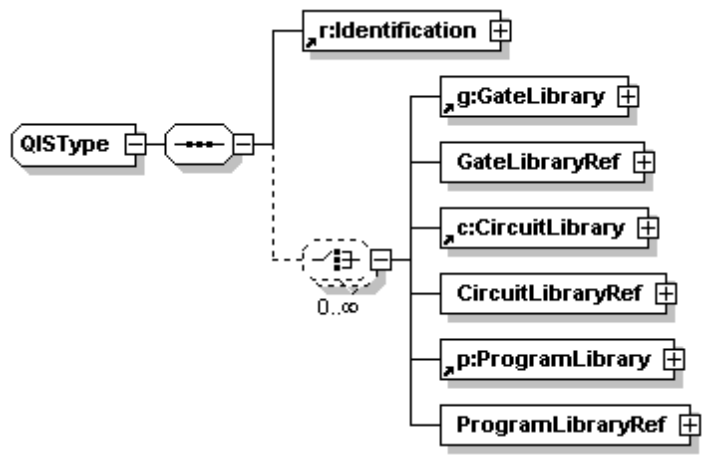

*Figure 1: The Instance Module*

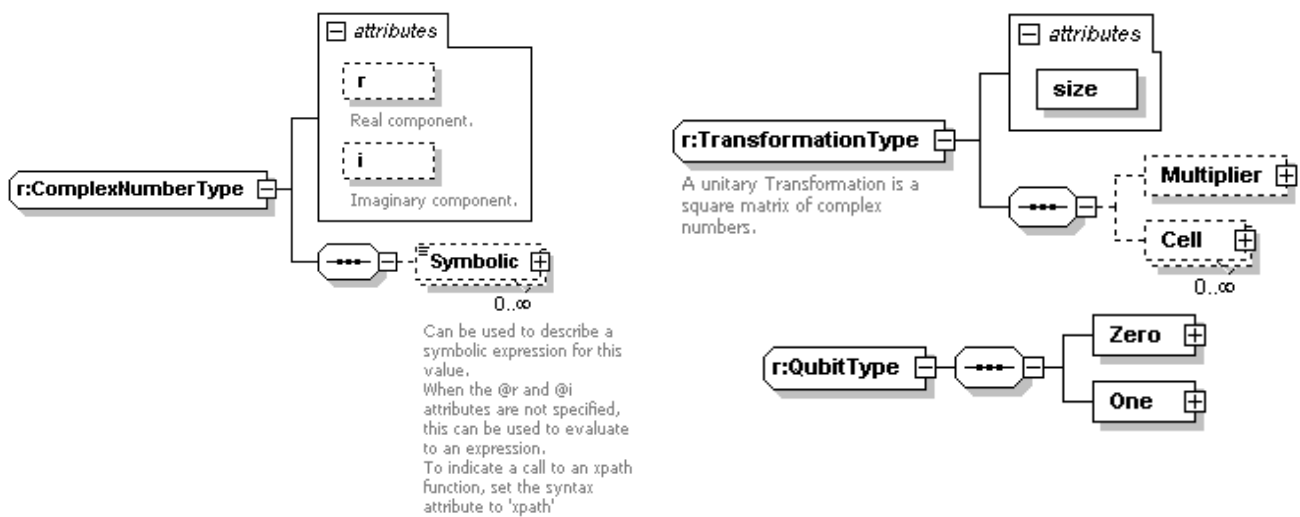

*Figure 2: Some fundamental Types*

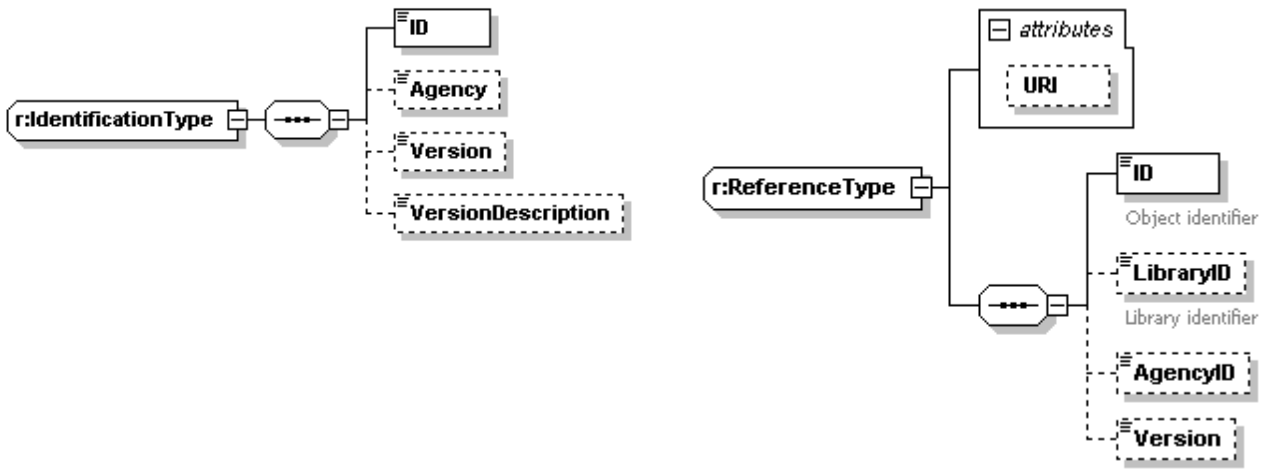

*Figure 3: The Identification and Reference types*

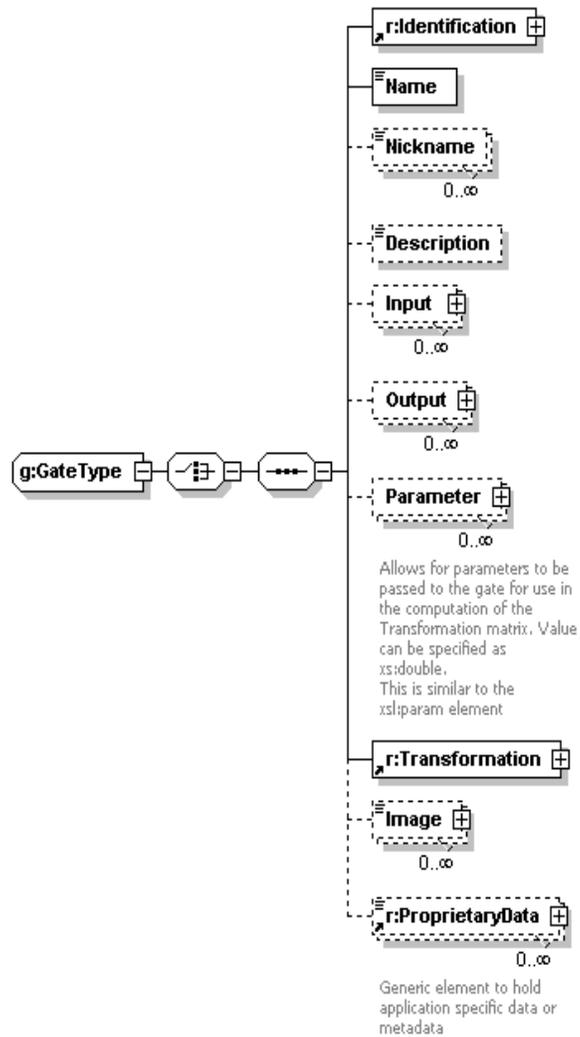

*Figure 4: The Gate Type*

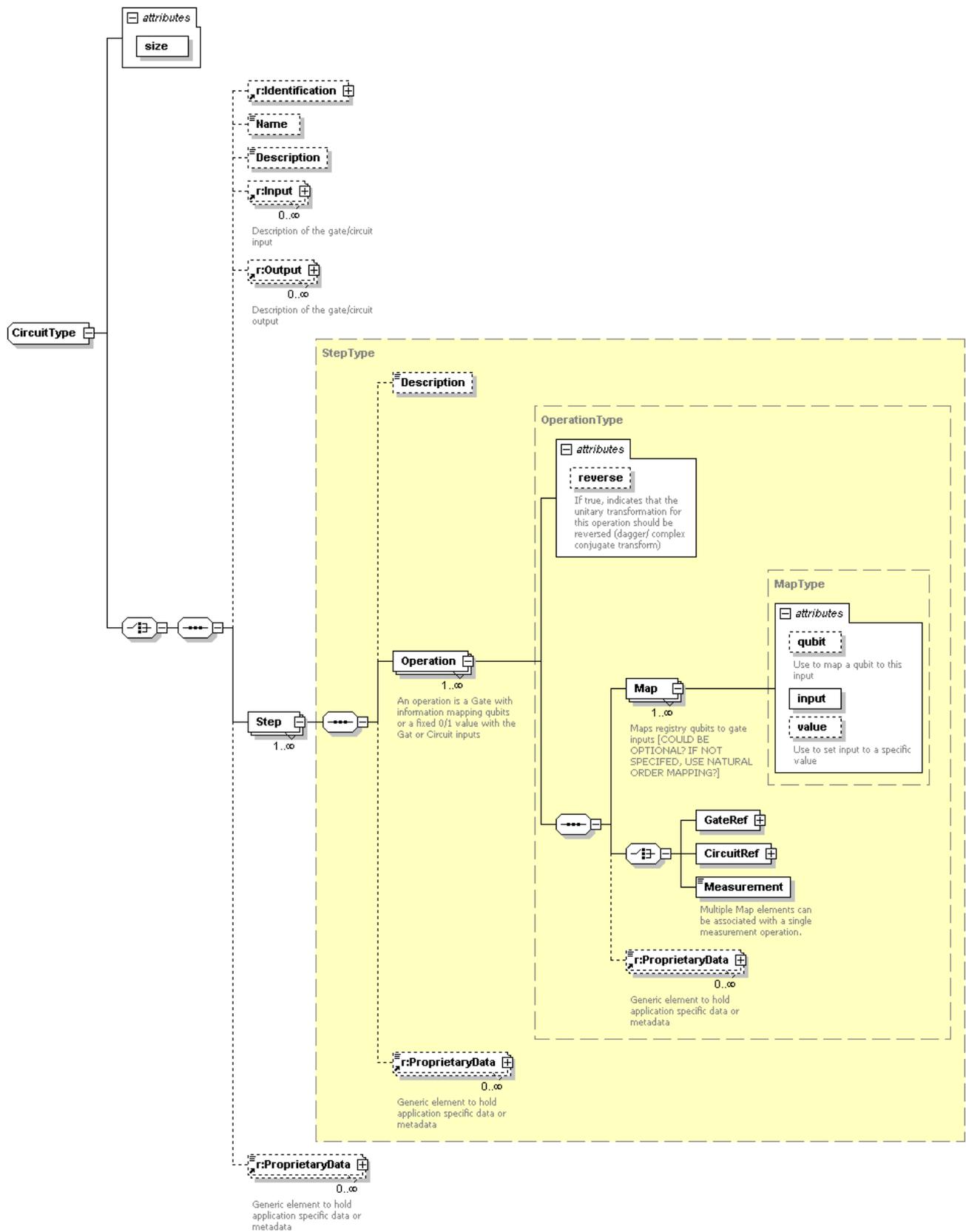

*Figure 5: The Circuit Type*

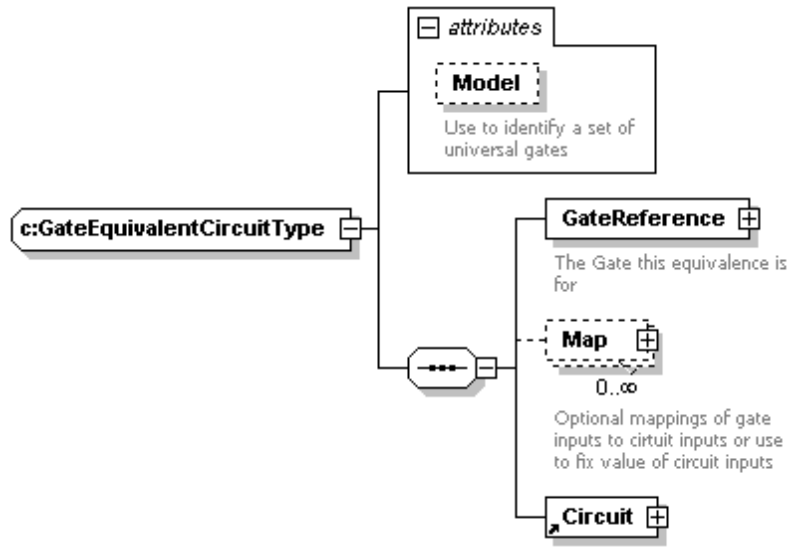

*Figure 6: The Gate Equivalence Circuit Type*

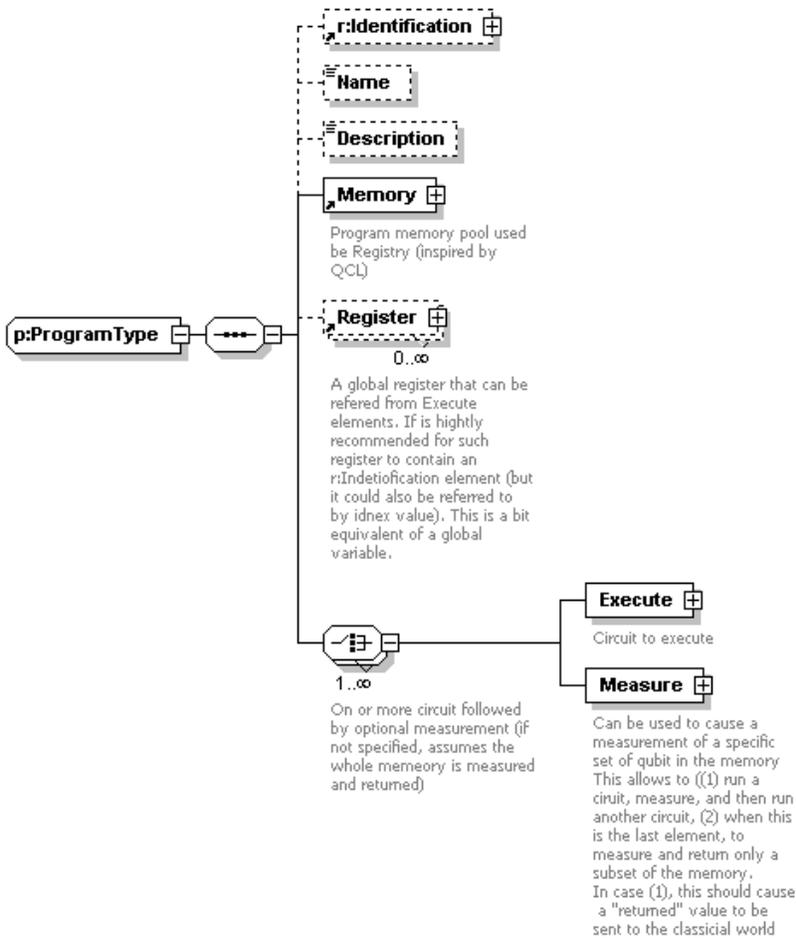

*Figure 7: The Program Type*

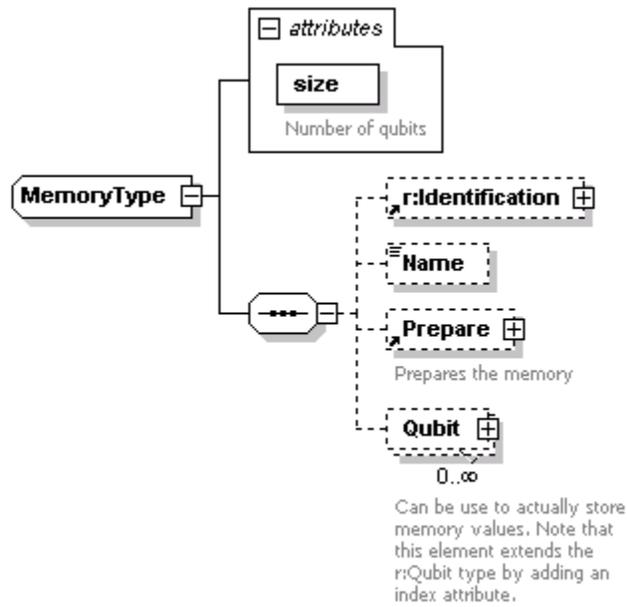

*Figure 8: The Memory Type*

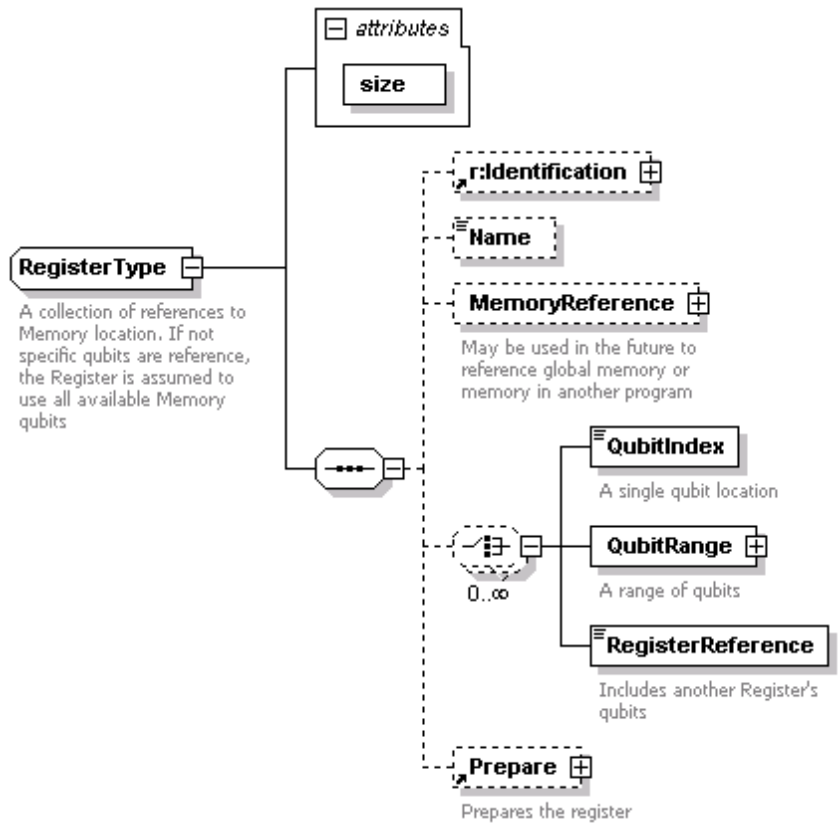

*Figure 9: The Register Type*

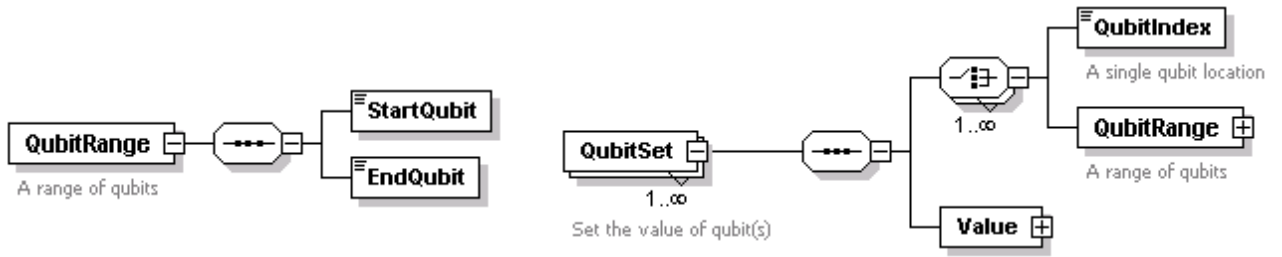

*Figure 10:Qubit Range and QubitSet types*

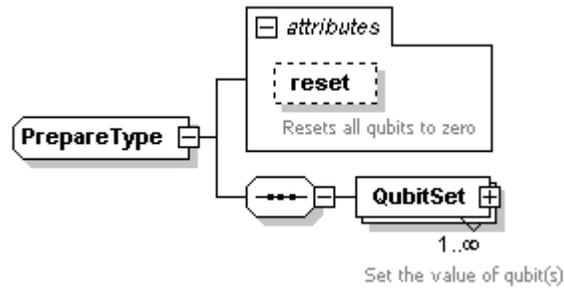

*Figure 11: The Prepare type*

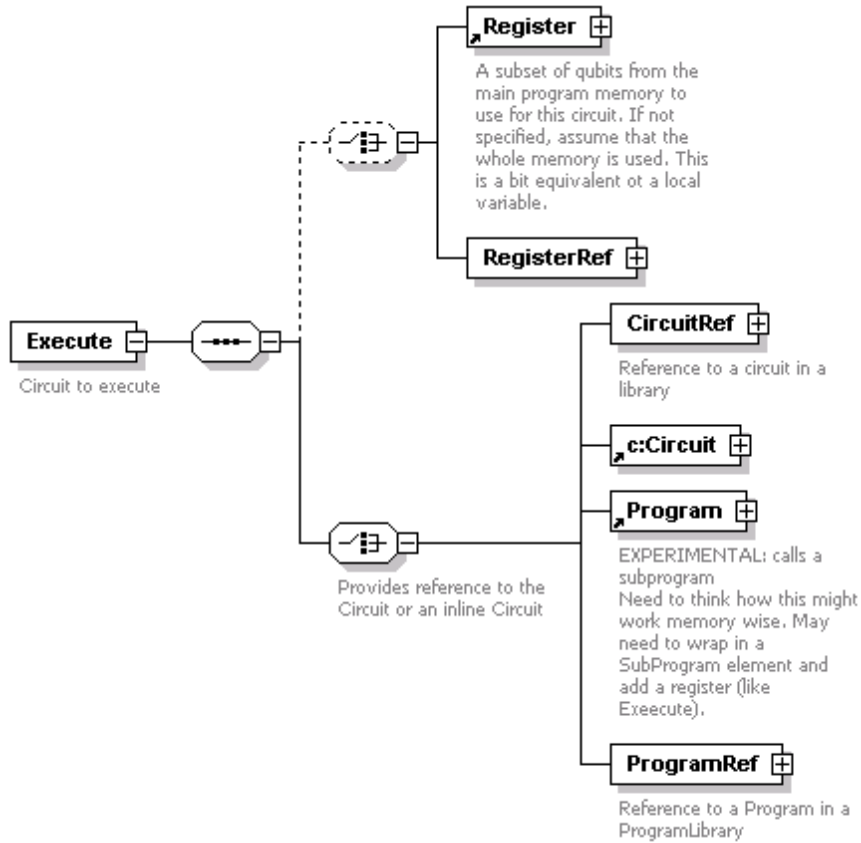

*Figure 12: The Execute Type*

# Annex 2: Output and visualization examples

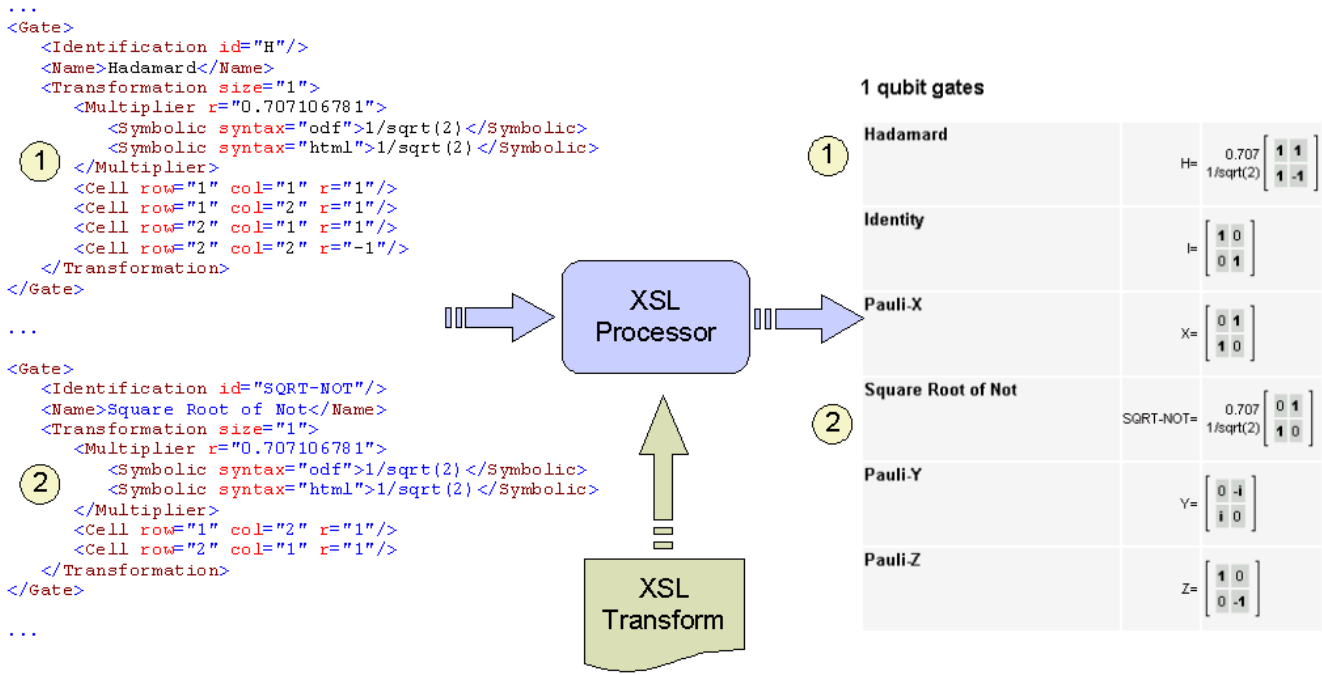

*Figure 13: Transformation of an XML document into HTML using XSLT and an XSL Processor*

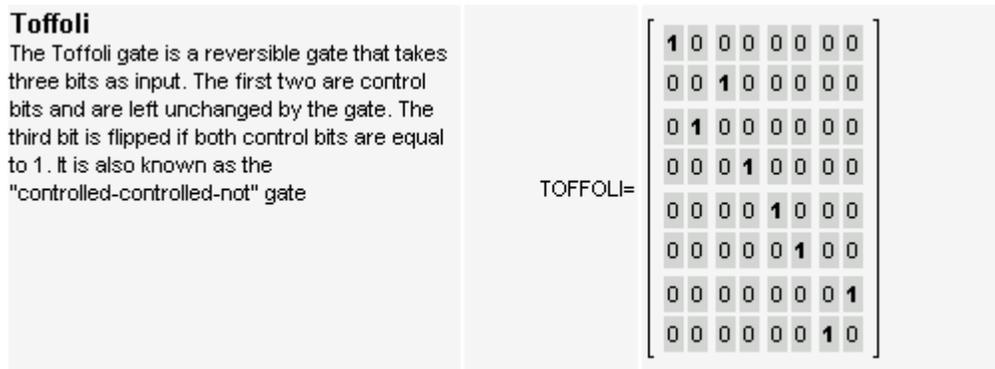

*Figure 14: An HTML representation of the Toffoli Gate*

```
Circuit id0x0606e4d8, Size 9, 5 step(s)
9-qubit Shor qubit code
Encoding circuit for the Shor nine qubit code.
  Step 1, 1 operation(s)
    Warning: Not all qubits have been mapped.
    1: Controlled-NOT (C-NOT) [1=1,4=3]
        ERROR: Map 1 input=3 is out of Gate range.
  Step 2, 1 operation(s)
    Warning: Not all qubits have been mapped.
    1: Controlled-NOT (C-NOT) [1=1,7=2]
  Step 3, 3 operation(s)
    Warning: Not all qubits have been mapped.
    1: Hadamard (H) [1=1]
    2: Hadamard (H) [4=1]
    3: Hadamard (H) [7=1]
  Step 4, 3 operation(s)
    Warning: Not all qubits have been mapped.
    1: Controlled-NOT (C-NOT) [1=1,2=2]
    2: Controlled-NOT (C-NOT) [4=1,5=2]
    3: Controlled-NOT (C-NOT) [7=1,8=2]
  Step 5, 3 operation(s)
    Warning: Not all qubits have been mapped.
    1: Controlled-NOT (C-NOT) [1=1,3=2]
    2: Controlled-NOT (C-NOT) [4=1,6=2]
    3: Controlled-NOT (C-NOT) [7=1,9=2]
```

*Figure 15: XSLT based second level validation for a 9-qubit Shor code circuit (in which an error has been introduced).*

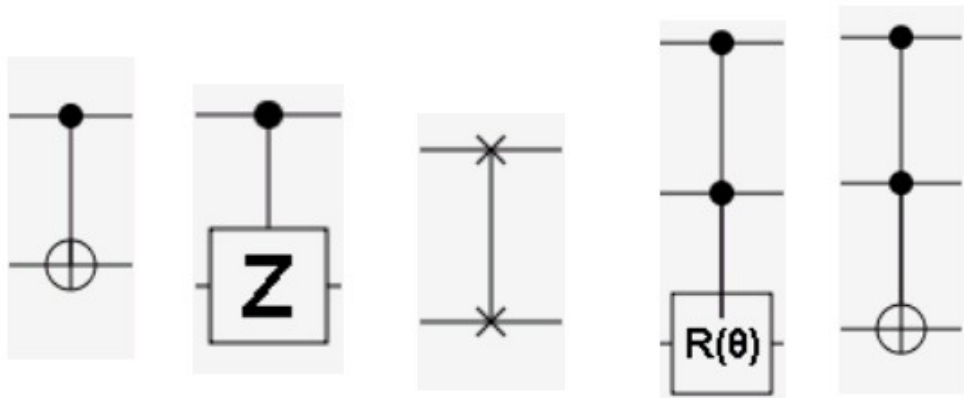

*Figure 16: SVG-XML representation of common quantum gates*

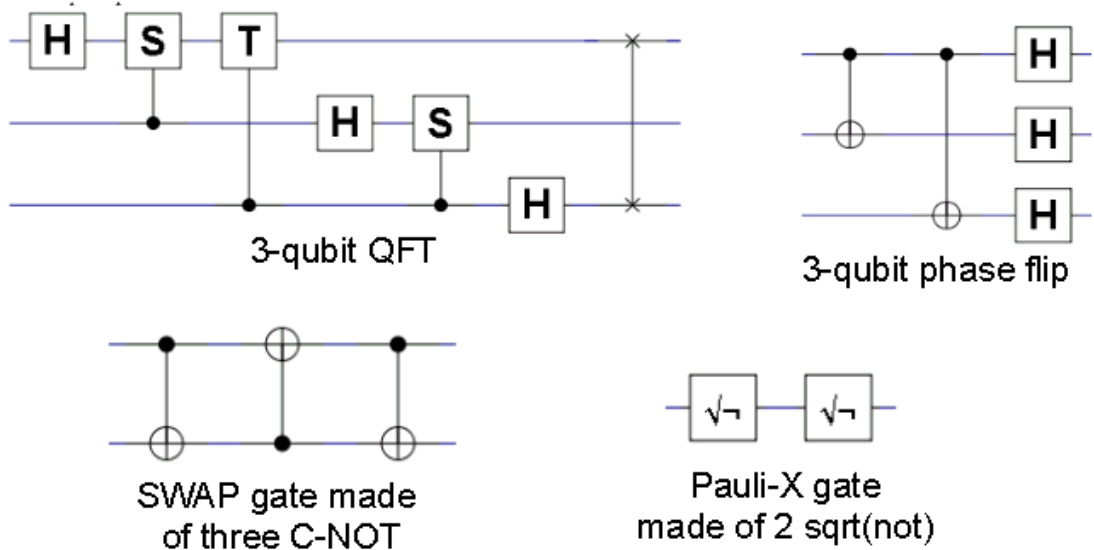

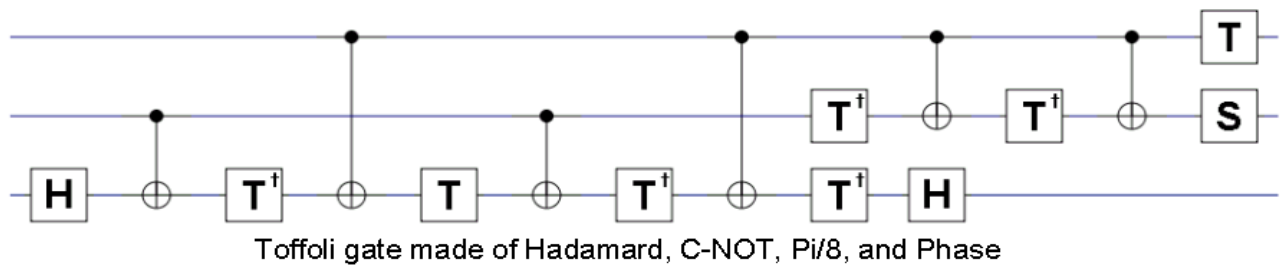

*Figure 17: SVG-XML representation of selected quantum circuits*

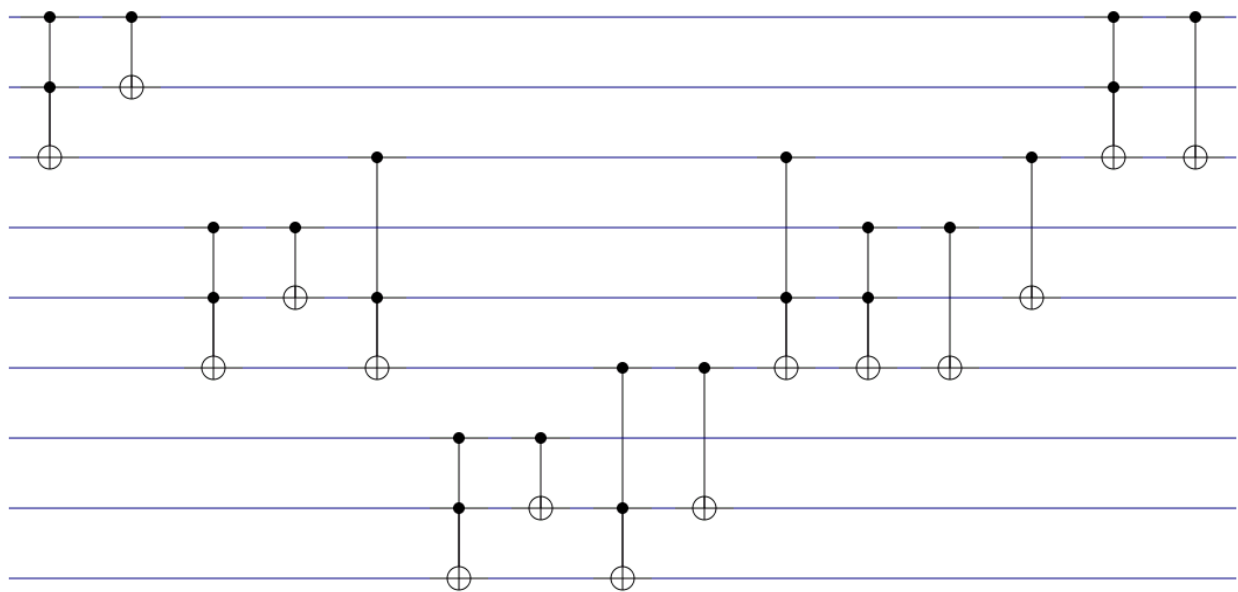

*Figure 18: SVG Version ofa a 3-qubit adder generated by the modified Genadder*

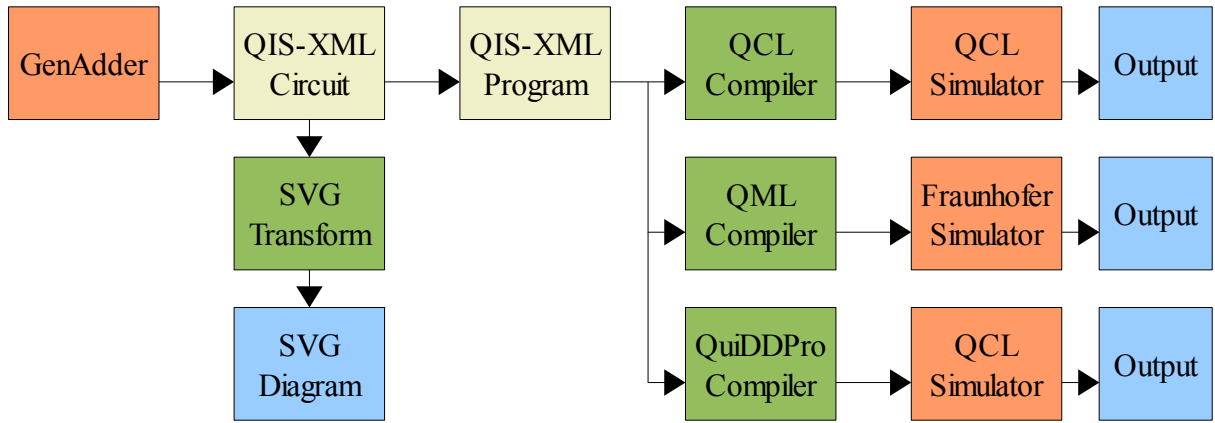

*Figure 19: Using QIS-XML to execute GenAdder circuits on various simulators*

```xml
<!-- 5 qubit adder 6+7 -->
<p:Program>
    <r:Identification>
        <r:ID>six_plus_seven</r:ID>
    </r:Identification>
    <p:Name>Six plus Seven</p:Name>
    <p:Memory size="15"/>
    <p:Execute>
        <p:Register size="15">
            <p:Prepare>
                <p:QubitSet>
                    <!-- A = 6 -->
                    <p:QubitIndex>4</p:QubitIndex>
                    <p:QubitIndex>7</p:QubitIndex>
                    <!-- B = 7 -->
                    <p:QubitIndex>2</p:QubitIndex>
                    <p:QubitIndex>5</p:QubitIndex>
                    <p:QubitIndex>8</p:QubitIndex>
                    <p:Value r="1"/>
                </p:QubitSet>
            </p:Prepare>
        </p:Register>
        <p:CircuitRef>
            <r:ID>adder5</r:ID>
        </p:CircuitRef>
    </p:Execute>
    <p:Measure>
        <p:Register size="6">
            <p:QubitIndex>2</p:QubitIndex>
            <p:QubitIndex>5</p:QubitIndex>
            <p:QubitIndex>8</p:QubitIndex>
            <p:QubitIndex>11</p:QubitIndex>
            <p:QubitIndex>14</p:QubitIndex>
            <p:QubitIndex>15</p:QubitIndex>
        </p:Register>
    </p:Measure>
</p:Program>
```

*Figure 20: QIS-XML Program to compute 6+7 using a 5-qubit adder*

```xml
<!-- ============================ -->
<!-- QIS-XML QML Compiler v2007.04 -->
<!-- ============================ -->
<QML>
        <Circuit Name="default" Size="6" Id="default.qml" Description="">
                <Operation Step="0"/>
                <Operation Step="1">
                        <Application Name="G" Id="0" Bits="1">
                                <Gate Type="PAULI_X"/>
                        </Application>
                        <Application Name="G" Id="0" Bits="3">
                                <Gate Type="PAULI_X"/>
                        </Application>
                </Operation>
                <Operation Step="2">
                        <Application Name="G" Id="0" Bits="0,1,2">
                                <Gate Type="TOFFOLI"/>
                        </Application>
                </Operation>
                <Operation Step="3">
                        <Application Name="G" Id="0" Bits="0,1">
                                <Gate Type="CNOT"/>
                        </Application>
                </Operation>
                <Operation Step="4">
                        <Application Name="G" Id="0" Bits="3,4,5">
                                <Gate Type="TOFFOLI"/>
                        </Application>
                </Operation>
                <Operation Step="5">
                        <Application Name="G" Id="0" Bits="3,4">
                                <Gate Type="CNOT"/>
                        </Application>
                </Operation>
                <Operation Step="6">
                        <Application Name="G" Id="0" Bits="2,4,5">
                                <Gate Type="TOFFOLI"/>
                        </Application>
                </Operation>
                <Operation Step="7">
                        <Application Name="G" Id="0" Bits="2,4">
                                <Gate Type="CNOT"/>
                        </Application>
                </Operation>
                <Operation Step="8">
                        <Application Name="G" Id="0" Bits="0,1,2">
                                <Gate Type="TOFFOLI"/>
                        </Application>
                </Operation>
                <Operation Step="9">
                        <Application Name="G" Id="0" Bits="0,2">
                                <Gate Type="CNOT"/>
                        </Application>
                </Operation>
        </Circuit>
</QML>
```

*Figure 21: QML generated code of a 2-qubit adder to perform 2+1*

```
// =============================
// QIS-XML QCL Compiler v2007.04
// =============================
int i;
int value;
// Allocate program memory
qureg memory[6];
// WARNING: p:Register not fully implemented ***
qureg registerIDAQ3MTE = memory;
// PREPARE
i = 1;
measure registerIDAQ3MTE[1],value;
if value != 1 { X(registerIDAQ3MTE[1]); }
measure registerIDAQ3MTE[3],value;
if value != 1 { X(registerIDAQ3MTE[3]); }

// CIRCUIT adder2
// STEP 1
// OPERATION 1
qureg registerIDAGFMTE = registerIDAQ3MTE[0]®isterIDAQ3MTE[1]®isterIDAQ3MTE[2];
CNot(registerIDAGFMTE[2] , registerIDAGFMTE[0] & registerIDAGFMTE[1]);

// STEP 2
// OPERATION 1
qureg registerIDATFMTE = registerIDAQ3MTE[0]®isterIDAQ3MTE[1];
CNot(registerIDATFMTE[1],registerIDATFMTE[0]);

// STEP 3
// OPERATION 1
qureg registerIDA3FMTE = registerIDAQ3MTE[3]®isterIDAQ3MTE[4]®isterIDAQ3MTE[5];
CNot(registerIDA3FMTE[2] , registerIDA3FMTE[0] & registerIDA3FMTE[1]);

// STEP 4
// OPERATION 1
qureg registerIDAKGMTE = registerIDAQ3MTE[3]®isterIDAQ3MTE[4];
CNot(registerIDAKGMTE[1],registerIDAKGMTE[0]);

// STEP 5
// OPERATION 1
qureg registerIDAUGMTE = registerIDAQ3MTE[2]®isterIDAQ3MTE[4]®isterIDAQ3MTE[5];
CNot(registerIDAUGMTE[2] , registerIDAUGMTE[0] & registerIDAUGMTE[1]);

// STEP 6
// OPERATION 1
qureg registerIDABHMTE = registerIDAQ3MTE[2]®isterIDAQ3MTE[4];
CNot(registerIDABHMTE[1],registerIDABHMTE[0]);

// STEP 7
// OPERATION 1
qureg registerIDALHMTE = registerIDAQ3MTE[0]®isterIDAQ3MTE[1]®isterIDAQ3MTE[2];
CNot(registerIDALHMTE[2] , registerIDALHMTE[0] & registerIDALHMTE[1]);

// STEP 8
// OPERATION 1
qureg registerIDAYHMTE = registerIDAQ3MTE[0]®isterIDAQ3MTE[2];
CNot(registerIDAYHMTE[1],registerIDAYHMTE[0]);

// MEASUREMENT
for i=0 to 5{
   measure memory[i],value;
   print i,"=",value;
}
```

*Figure 22: QCL generated program to compute 2+1*